\documentclass[twocolumn,aps,pre,showpacs]{revtex4}
\usepackage{amssymb, amsmath, epsfig}

\begin{document}

\title{Non-monotonous short-time decay of the Loschmidt echo\\ in
  quasi-one-dimensional systems}

\author{Arseni Goussev}

\affiliation{Institut f\"ur Theoretische Physik, Universit\"at
  Regensburg, 93040 Regensburg, Germany}

\affiliation{School of Mathematics, University of Bristol, University
  Walk, Bristol BS8 1TW, United Kingdom}

\date{\today}

\begin{abstract}
  We study the short-time stability of quantum dynamics in
  quasi-one-dimensional systems with respect to small localized
  perturbations of the potential. To this end, we address,
  analytically and numerically, the decay of the Loschmidt echo (LE)
  during times short compared to the Ehrenfest time. We find that the
  LE is generally a non-monotonous function of time and exhibits
  strongly pronounced minima and maxima at the instants of time when
  the corresponding classical particle traverses the perturbation
  region. We also show that, under general conditions, the envelope
  decay of the LE is well approximated by a Gaussian, and we derive
  explicit analytical formulas for the corresponding decay
  time. Finally, we demonstrate that the observed non-monotonicity of
  the LE decay is only pertinent to one-dimensional (and, more
  generally, quasi-one-dimensional systems), and that the short-time
  decay of the LE can be monotonous in higher number of dimensions.
\end{abstract}


\maketitle

\section{Introduction}
\label{section_introduction}

Nearly a quarter of a century has passed since Peres
\cite{Per84Stability} made an important advance in understanding the
origin of irreversibility in quantum theory. He pioneered a framework
in which dynamical instabilities were studied in terms of small
external perturbations of the Hamilton operator without resorting to
the classical concepts of mixing and coarse graining. In particular,
Peres convincingly demonstrated that a change in the time evolution of
a quantum system caused by the perturbation is largely determined by
whether the system exhibits regular or chaotic behavior in the
classical limit.

The quantity addressed by Peres, currently known as the {\it Loschmidt
  echo} (LE) in the quantum chaos community and the {\it fidelity} in
the field of quantum information, is defined as \cite{Per84Stability}
\begin{equation}
  M(t) = \left| \langle \Psi_0 | \exp(i\hat{H}_2 t/\hbar)
    \exp(-i\hat{H}_1 t/\hbar) | \Psi_0 \rangle \right|^2 .
\label{le-def}
\end{equation}
It is the squared overlap of the initial state $| \Psi_0 \rangle$ and
the state obtained by propagating $| \Psi_0 \rangle$ though time $t$
under the original, unperturbed Hamiltonian $\hat{H}_1$, and then
through time $-t$ under the perturbed Hamiltonian $\hat{H}_2$. In
other words, the LE, $M(t)$, is a measure of the time-reversibility of
a system's dynamics under ``imperfect conditions''. In an alternative
interpretation, the LE characterizes the distance between two states
in the Hilbert space, $| \Psi_1 \rangle = \exp(-i\hat{H}_1 t/\hbar) |
\Psi_0 \rangle$ and $| \Psi_2 \rangle = \exp(-i\hat{H}_2 t/\hbar) |
\Psi_0 \rangle$, both obtained from the same initial state, $| \Psi_0
\rangle$, in the course of the quantum evolution through time $t$
under two different Hamiltonians, $\hat{H}_1$ and $\hat{H}_2$
respectively. By construction, $M(0) = 1$ assuming the initial state
is normalized. Typically, $M(t)$ decreases (or ``decays'') with
increasing time, and the precise functional form of the decay is
determined by the nature of the Hamiltonians $\hat{H}_1$ and
$\hat{H}_2$, as well as by the initial state.

Interest to the subject of the LE revived a decade ago. This was
largely due to the discovery of a perturbation independent regime
\cite{JP01Environment} of the LE decay in systems that exhibit chaotic
dynamics in the classical limit. In this regime, known as the Lyapunov
regime, the LE decays exponentially in time, $M(t) \sim \exp(-\lambda
t)$, with the decay rate $\lambda$ equal to the average Lyapunov
exponent of the corresponding classical system. The Lyapunov regime
can be considered as a clear example of how classical chaos manifests
itself in a quantum mechanical world.

Roughly speaking, for chaotic systems under the action of {\it global}
Hamiltonian perturbations, i.e., perturbations affecting a dominant
part of the system's phase space, one identifies three distinct decay
regimes of the LE: the Gaussian perturbative regime \cite{JSB01Golden,
  CT02Sensitivity} for ``weak'' perturbations, the exponential
Fermi-golden-rule (FGR) regime \cite{JP01Environment, JSB01Golden,
  CT02Sensitivity, Pro02General} for ``moderate'' perturbation
strengths, and the exponential Lyapunov regime \cite{JP01Environment,
  CLM+02Measuring} for ``strong'' perturbations. The above three
regimes serve as a basis for the theory of the LE decay for global
perturbation. The full theory however is much more subtle and the
interested reader is referred to two comprehensive review articles,
Refs.~\cite{GPSZ06Dynamics, JP09Decoherence}.

The LE due to global perturbations has long been a subject of numerous
experimental studies. Most notably these are experiments in nuclear
magnetic resonance \cite{Hah50Spin, ZME92Polarization, UPL98Gaussian,
  PLU+00nuclear}, quantum optics \cite{KAH64Observation}, cold atoms
\cite{BDL+00Wave, AKD03Echo, AGKD04Revivals, AKGD06Decay,
  WTP09Observation}, superconductivity \cite{NPYT02Charge}, microwave
cavities \cite{SSGS05Experimental, SGSS05Fidelity, BZK+09Probing,
  KKS+10Microwave}, and elastodynamics \cite{GSW06Scattering,
  LW08Scattering}.

In recent years, Hamiltonian perturbations of a new kind, namely
perturbations that are {\it local} in phase space, have been addressed
in the context of the LE decay both theoretically and experimentally
\cite{GR07Loschmidt, HKS08Algebraic, GWRJ08Loschmidt, AW09Loschmidt,
  KKS+11Fidelity}. In a semiclassical picture, a local perturbation is
concentrated in a small region of the system's phase space, so that
the length of a typical trajectory between any two successive visits
of the perturbation region is much larger than the system's size. In
strongly chaotic systems the phase space extent of a local
perturbation can be characterized by an ``escape'' rate
$\tau_{\mathrm{esc}}^{-1}$ defined as the rate at which trajectories
of the corresponding classical system visit the perturbation
region. The semiclassical analysis of the LE for times longer than the
Ehrenfest time, i.e., longer than the time it takes for an initially
localized wave packet to explore the available phase space, has
revealed an exponential decay regime, $M(t) \sim \exp(-\kappa t)$, in
which the decay rate $\kappa$ is a non-monotonous function of the
perturbation strength \cite{GR07Loschmidt, GWRJ08Loschmidt}. In
particular, for sufficiently weak perturbations the FGR regime is
recovered in which $\kappa$ grows quadratically with the perturbations
strength. However, in the limit of strong perturbations $\kappa$
saturates at a perturbation independent value
$2\tau_{\mathrm{esc}}^{-1}$ corresponding to the so-called escape-rate
regime. The crossover from the FGR to the escape-rate regime is
non-monotonous, and $\kappa$ exhibits well pronounced oscillations as
a function of the perturbations strength. These oscillations have been
then confirmed in numerical experiments with perturbed cat maps
\cite{AW09Loschmidt} and, more recently, in laboratory experiments
with microwave cavities \cite{KKS+11Fidelity}. Finally, the limit of
point-like perturbations, for which $\tau_{\mathrm{esc}} \rightarrow
\infty$, was addressed in Ref.~\cite{HKS08Algebraic}. There, the LE
was shown to decay algebraically with time, $M(t) \sim t^{-2}$.

The existing theory of the LE decay from local perturbations is only
applicable to times long compared to the Ehrenfest time. However, in
certain cases the short time decay of the echo, during which the
unperturbed and perturbed quantum states can be described by localized
wave packets, might be of significant importance for understanding
results of laboratory experiments. For instance, in echo spectroscopy
experiments with cold atoms trapped inside an optical billiard
\cite{AKD03Echo, AGKD04Revivals, AKGD06Decay} one typically observes
the echo decay for times as short as only few free flight times of the
corresponding classical billiard. Such time scales can be short
compared to the Ehrenfest time depending on a parameter choice. It is
the objective of this paper to address the short-time decay of the LE
due to local Hamiltonian perturbations. More specifically, we make the
first step in this direction by performing a detailed study of the LE
in one-dimensional (and, more generally, quasi-one-dimensional)
systems in the presence of localized perturbations of the system's
potential. We show that in such systems $M(t)$ is typically a
non-monotonous function exhibiting well pronounced minima and maxima
at times when the particle traverses the perturbation region. We also
show that in closed systems the envelope of $M(t)$ can be well
approximated by a Gaussian, $\exp\left[ -(t/\tau)^2 \right]$, with the
decay time $\tau$ explicitly expressible in terms of system
parameters. All results presented in this paper concern the LE decay
in ``clean'' systems and imply no averaging over initial states or
Hamiltonian perturbations.

The analysis presented in this paper only concerns the short-time
decay of the LE in conservative quasi-one-dimensional, and therefore
essentially integrable, systems. The case of the LE decay in
classically chaotic systems for times shorter than the Ehrenfest time
(and in the presence of global perturbations) was addressed in
Ref.~\cite{STB03Hypersensitivity}. There, for a ``typical'' localized
initial state $\Psi_0$, the LE was shown to exhibit the
double-exponential initial decay, $M(t) \sim \exp (-\mathrm{constant}
\times e^{2\lambda t})$ with $\lambda$ being the Lyapunov exponent.

The paper is organized as follows. In Section~\ref{section_1d} we
provide the analysis of the LE decay in one-dimensional systems based
on full numerical solution of the time-dependent Schr\"odinger
equation and on the thawed Gaussian approximation (TGA) in its
standard and modified, average potential formulations. Details on the
average potential TGA are deferred to
Appendix~\ref{section_TGA}. Particular examples treated in
Section~\ref{section_1d} include a free particle
(Sec.~\ref{section_free}), particle on a ring
(Sec.~\ref{section_ring}), harmonic (Sec.~\ref{section_harmonic}) and
anharmonic (Sec.~\ref{section_anharmonic})
oscillators. Section~\ref{section_2d} demonstrates disappearance of
non-monotonous features of the LE decay as a quasi-one-dimensional
system is transformed into a substantially two-dimensional system. In
Section~\ref{section_conclusions} we give a discussion of our results
and make concluding remarks.

\section{Loschmidt echo in one dimension}
\label{section_1d}

\subsection{Free particle}
\label{section_free}

\begin{figure}[h]
\centerline{\epsfig{figure=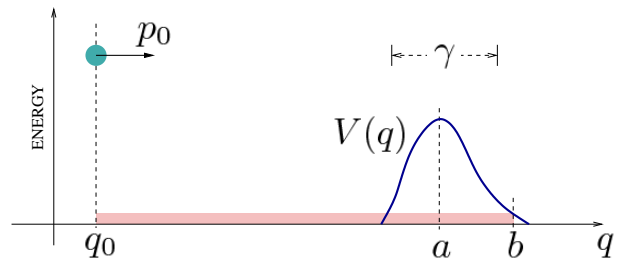,width=3.2in}}
\caption{(Color online) One-dimensional quantum particle in the
  presence of a localized potential barrier. See the text for
  discussion.}
\label{fig-1}
\end{figure}

The essential physics of the short-time decay of the LE for local
potential perturbations can already be revealed by considering the
simplest physical system -- a free particle. In this scenario, the
unperturbed Hamiltonian, $\hat{H}_1 = \hat{p}^2/2m$, describes
one-dimensional motion of a free particle of mass $m$; in the
coordinate representation, the momentum operator $\hat{p} =
-i\hbar \partial/\partial q$. The perturbed Hamiltonian, $\hat{H}_2 =
\hat{H}_1 + V(\hat{q})$, represents a particle interacting with a
potential ``barrier'' $V(q)$ localized to an interval on the position
axis, see Fig.~\ref{fig-1}. For concreteness, we assume $V(q)$ to have
a single maximum (or minimum) at $q=a$. (As it will become clear from
the following presentation, our results can be easily generalized to
perturbation potentials of arbitrary shape.) We also label the
characteristic extent of the potential by $\gamma$, see
Fig.~\ref{fig-1}. Finally, we set the initial state to be a Gaussian
wave packet,
\begin{equation}
  \Psi_0(q) = \left( \frac{1}{\pi \sigma^2} \right)^{\frac{1}{4}} 
  \exp\left( \frac{i}{\hbar} p_0(q-q_0) - \frac{(q-q_0)^2}{2\sigma^2} \right)
\label{init_wp}
\end{equation}
that represents a quantum particle with the average position
coordinate $q_0$ and momentum $p_0$. The initial wave packet
dispersion is quantified by $\sigma$. We further assume that the
initial state, $\Psi_0(q)$, has zero overlap with the perturbation
potential, i.e.,
\begin{equation}
  \gamma,\sigma \ll l \, ,
\label{condition-1}
\end{equation}
where $l = |a - q_0|$ is the characteristic distance between the
particle and the perturbation region, see Fig.~\ref{fig-1}. We also
assume that the initial state is well localized in the momentum space,
so that
\begin{equation}
  p_0 \sigma \gg \hbar \, ,
\label{condition-2}
\end{equation}
and that the potential $V(q)$ constitutes a classically small
perturbation:
\begin{equation}
  |V(q)| \ll E \, ,
\label{condition-3}
\end{equation}
where $E = p_0^2/2m$ is the total energy of the corresponding
classical particle. Together Eqs.~(\ref{condition-2}) and
(\ref{condition-3}) imply that the probability density gets almost
perfectly transmitted over the barrier in the course of the time
evolution, and the effect of quantum reflections can be neglected.

Most of semiclassical studies of the LE decay rely on the so-called
``dephasing representation'' (DR) \cite{Vanicek04,Vanicek06}, in which
the LE amplitude $\langle \Psi_2(t) | \Psi_1(t) \rangle$ is expressed
as an interference integral over trajectories of the unperturbed
system, modulated by a ``dephasing'' factor due to the perturbation,
with initial phase space coordinates weighed by the Wigner function of
the state $| \Psi_0 \rangle$.  Our analysis however is based on a
different analytical approach, namely on the ``thawed Gaussian''
approximation (TGA) \cite{Hel75Time,Hel06Guided,Tan07Introduction} in
its standard and extended versions, see Appendix~\ref{section_TGA} and
the discussion below. The TGA, unlike the DR, involves only a single
trajectory transporting the wave packet center through phase space in
the course of its time evolution. This allows for greater analytical
flexibility at the expense of a reduction in accuracy. For most of our
purposes however the TGA proves to be sufficiently reliable both
qualitatively and quantitatively.

According to the TGA both the unperturbed and perturbed wave functions
preserve their Gaussian form in the course of their time evolution:
\begin{align}
  \Psi_j&(q;t) = \left( \frac{2 \Re \alpha_j}{\pi} \right)^{\frac{1}{4}} \nonumber\\
  & \times \exp\left( -\alpha_j (q-q_j)^2 + \frac{i}{\hbar} p_j
    (q-q_j) + \frac{i}{\hbar} \phi_j \right) ,
\label{wp_time}
\end{align}
where $j=1,2$ labels the unperturbed and perturbed wave functions
respectively, and $\Re$ stands for the real part. Here, $q_j(t)$,
$p_j(t)$, $\phi_j(t)$ are real-valued and $\alpha_j(t)$ complex-valued
functions of time that parametrize the wave functions. In the
``standard'' TGA \cite{Hel75Time,Hel06Guided,Tan07Introduction} the
time evolution of these parameters is governed by
\begin{align}
  &\dot{q}_j = p_j/m \, , \label{tga.a} \\
  &\dot{p}_j = -W_j'(q_j) \, , \label{tga.b} \\
  &\dot{\alpha}_j = -\frac{2i\hbar}{m}\alpha_j^2 + \frac{i}{2\hbar} W_j''(q_j) \, , \label{tga.c}
\end{align}
and $\dot{\phi}_j = p_j^2/2m - W_j(q_j) -
(\hbar^2/m)\Re\alpha_j$. Here $W_j(q)$ is the corresponding
(unperturbed for $j=1$ and perturbed for $j=2$) potential, and the
prime denotes differentiation with respect to $q$. The system of four
real, first order ordinary differential equations given by
Eqs.~(\ref{tga.a}--\ref{tga.c}) is then solved with the initial
conditions $q_j = q_0$, $p_j = p_0$, and $\alpha_j^{-1} = 2\sigma^2$
at $t = 0$. In the ``average potential'' TGA, see
Appendix~\ref{section_TGA}, Eqs.~(\ref{tga.a}--\ref{tga.c}) are
replaced by
\begin{align}
  &\dot{q}_j = p_j/m \, , \label{avr_tga.a} \\
  &\dot{p}_j = -\widetilde{W}_j'(q_j;\Re\alpha_j) \, , \label{avr_tga.b} \\
  &\dot{\alpha}_j = -\frac{2i\hbar}{m}\alpha_j^2 + \frac{i}{2\hbar} \widetilde{W}_j''(q_j;\Re\alpha_j) \, , \label{avr_tga.c}
\end{align}
with the average potential $\widetilde{W}_j$ defined as
\begin{equation}
  \widetilde{W}_j(q;s) = (2s/\pi)^{\frac{1}{2}} \int dx \, W_j(x) e^{ -2s (x-q)^2 } .
\label{avr_pot}
\end{equation}
As before the prime denotes partial differentiation with respect to
$q$.

In the free particle case the unperturbed wave function $\Psi_1(q;t)$
given by the TGA, Eq.~(\ref{wp_time}), is exact and the parameters
$q_1$, $p_1$, and $\alpha_1$ evolve in time according to
\cite{Tan07Introduction}
\begin{align}
  &q_1 = q_0 + p_0 t/m \, , \label{wp_1.a} \\
  &p_1 = p_0 \, , \label{wp_1.b} \\
  &\alpha_1^{-1} = 2 (\sigma^2 + i \hbar t / m) \, ,\label{wp_1.c}
\end{align}
and $\phi_1 = (p_0^2 / 2m) \, t - (\hbar/2) \arctan(\hbar t / m
\sigma^2)$.

Time evolution of the parameters $q_2$, $p_2$, $\alpha_2$
characterizing the perturbed wave function $\Psi_2(q;t)$ are
determined from a set of three ordinary differential equations
(\ref{tga.a}--\ref{tga.c}) in the ``standard TGA'' and by equations
(\ref{avr_tga.a}--\ref{avr_pot}) in the ``average potential TGA'' with
$j=2$ and $W_2(q) = V(q)$. Here we note that the LE defined by
Eq.~(\ref{le-def}) is unaffected by the global phases $\phi_1$ and
$\phi_2$ in the regime that allows one to approximate the unperturbed
and perturbed states by the simple Gaussian wave packets given by
Eq.~(\ref{wp_time}). The phase $\phi$ of an individual Gaussian wave
packet would only be physically relevant if the initial state $\Psi_0$
was a superposition of two or more Gaussian wave packets, or if one
was interested in the LE amplitude $\langle \Psi_2 (t) | \Psi_1 (t)
\rangle$ rather than in $M(t)$. However, considerations of these kinds
go beyond the scope of the current study.

Substituting the two time-dependent Gaussian wave packets, given by
Eq.~(\ref{wp_time}) with $j=1$ and $2$, into Eq.~(\ref{le-def}) we
obtain for the LE
\begin{align}
  M(t) = &\frac{2 \sqrt{\Re\alpha_1 \Re\alpha_2}}{|\alpha_1 +
    \alpha_2^*|} \exp \Bigg[ -\frac{2}{|\alpha_1 + \alpha_2^*|^2}
  \nonumber\\ &\times \bigg( \Re\big( \alpha_1\alpha_2
  (\alpha_1+\alpha_2)^* \big) \Delta q^2 \nonumber\\ &+ \Im(\alpha_1
  \alpha_2) \frac{\Delta q \Delta p}{\hbar} + \Re(\alpha_1 + \alpha_2)
  \frac{\Delta p^2}{4\hbar^2} \bigg) \Bigg] \, ,
\label{le_general}
\end{align}
where $\Delta q = q_2 - q_1$ and $\Delta p = p_2 - p_1$, asterisk
denotes complex conjugation, and $\Im$ stands for the imaginary
part. Equation~(\ref{le_general}), along with equations describing
time evolution of the parameters $q_j$, $p_j$, and $\alpha_j$ with
$j=1,2$, provides the main framework for our analytical study of the
LE.

We also perform a numerical analysis of the problem of the LE
decay. To this end we adopt a method of expending the quantum
propagator $\exp(-i\hat{H}t/\hbar)$ in terms of Chebyshev polynomials
of the Hamiltonian $\hat{H}$. For a detailed discussion of the method
see Ref.~\cite{RKMF03Unified} and references within. Hereinafter we
refer to results of the numerical solution of the LE decay problem as
to ``exact'' results as opposed to approximate ones obtained by using
the standard and average potential TGA.

\begin{figure}[h]
\centerline{\epsfig{figure=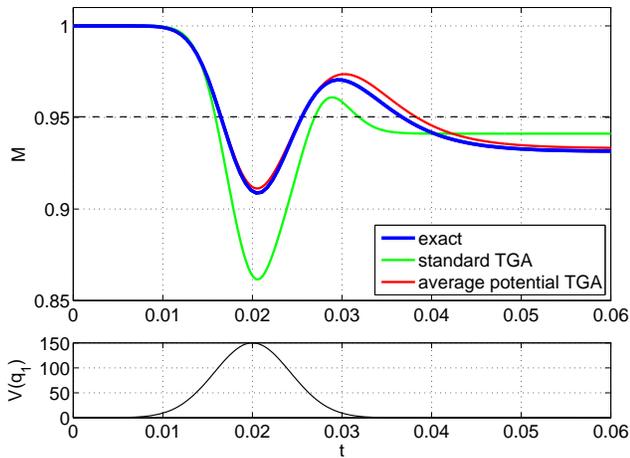,width=3.3in}}
\caption{(Color online) Top figure: The LE, $M(t)$, for a free
  particle with the perturbation potential $V(q)$ given by
  Eq.~(\ref{V_Gauss}). The horizontal dashed line represents the LE
  saturation value $M(\infty)$ given by Eq.~(\ref{le_satur_2}). Bottom
  figure: Perturbation potential as seen by the unperturbed classical
  particle, i.e. $V(q_1(t))$. The system is characterized by $p_0 =
  50$, $\sigma = 0.1$, $a-q_0 = 1$, $\gamma = 0.3$, and $V_0 = 150$.}
\label{fig-2}
\end{figure}

Figure~\ref{fig-2} shows the LE decay for the case of a perturbation
potential given by the Gaussian barrier
\begin{equation}
  V(q) = V_0 \exp\left( -\frac{(q-a)^2}{\gamma^2} \right) .
\label{V_Gauss}
\end{equation}
The upper part of the figure displays $M(t)$. Here, the thick blue
curve shows a result of the full, quantum-mechanical computation of
the LE defined by Eq.~(\ref{le-def}), i.e., the exact LE decay. The
green curve is obtained by using the standard TGA, i.e.,
Eq.~(\ref{le_general}) with the time evolution of parameters $q_2$,
$p_2$, and $\alpha_2$ determined from
Eqs.~(\ref{tga.a}--\ref{tga.c}). The red curve represents the LE decay
obtained in the average potential TGA by using
Eqs.~(\ref{avr_tga.a}--\ref{avr_pot}) for the time evolution of $q_2$,
$p_2$, and $\alpha_2$. The initial wave packet, Eq.~(\ref{init_wp}),
is specified by $q_0 = -1$, $p_0 = 50$, and $\sigma = 0.1$
\footnote{Dimensionless (atomic) units, $\hbar = m = 1$, are used in
  all numerical examples throughout the paper.}. The perturbation
potential is given by Eq.~(\ref{V_Gauss}) with $a = 0$, $\gamma =
0.3$, and $V_0 = 150$. The lower part of Fig.~\ref{fig-2} shows the
function $V(q_1(t))$, which has the meaning of the perturbation
potential as ``seen'' by the unperturbed classical particle.

Figure~\ref{fig-2} prompts an interesting observation: the LE develops
a minimum at the time the unperturbed classical particle passes the
extremum of the perturbation potential, and a maximum at the time the
particle exits the perturbation region. In fact, such behavior is
generic. This can be shown by imposing the approximation $\alpha_2(t)
\simeq \alpha_1(t)$, which simplifies Eq.~(\ref{le_general}) to
\begin{equation}
  M(t) = \exp \left[ -\frac{1}{\Re\alpha_1} \left( |\alpha_1|^2
  \Delta q^2 + \Im\alpha_1 \frac{\Delta q \Delta p}{\hbar} +
  \frac{\Delta p^2}{4\hbar^2} \right) \right] \, .
\label{le_same_alpha}
\end{equation}
Then, substituting Eq.~(\ref{wp_1.c}) into Eq.~(\ref{le_same_alpha})
we obtain
\begin{equation}
  M(t) = \exp \left[ -\frac{1}{2\sigma^2} \left( \Delta q - \frac{t}{m} \Delta p \right)^2
    - \frac{1}{2} \left( \frac{\sigma \Delta p}{\hbar} \right)^2 \right] \, .
\label{le_free}
\end{equation}
The function $M(t)$, given by Eq.~(\ref{le_free}), possesses two
extrema at time instants determined by
\begin{align}
  &\displaystyle \frac{d}{dt} \Delta p = 0 \quad &\mathrm{(minimum)} \, , \label{min_max.a} \\
  &\Delta q = \left[ 1 + \left( \frac{m \sigma^2}{\hbar t} \right)^2
  \right] \frac{t}{m}\Delta p \quad &\mathrm{(maximum)} \,
  . \label{min_max.b}
\end{align}
We now utilize the standard TGA, Eqs.~(\ref{tga.a}) and (\ref{tga.b})
with $j=2$ and $W_2(q) = V(q)$, to express the position and momentum
shifts, $\Delta q$ and $\Delta p$, in terms of the perturbation
$V(q)$. To the leading order in $V_0/E$, with $V_0 = \max_q |V(q)|$,
we obtain
\begin{align}
  &\Delta q(t) = -\frac{1}{p_0} \int_{0}^{t} dt'
  V(q_1(t')) \, , \label{dq_dp.a} \\
  &\Delta p(t) = -\frac{m}{p_0} V(q_1(t)) \, . \label{dq_dp.b}
\end{align}
Substitution of Eq.~(\ref{dq_dp.b}) into Eq.~(\ref{min_max.a}) shows
that $M(t)$ exhibits a minimum at time $t_a = m (a-q_0)/ p_0 =
ml/|p_0|$, with $a$ being the extremum point of the perturbation
potential, see Fig.~\ref{fig-1}. It is the time instant at which the
unperturbed classical particle passes the extremum point of the
perturbation potential, i.e., $q_1(t_a) = a$. Similarly, the LE
exhibits a maximum at time $t_b = m (b-q_0)/ p_0$ that, in accordance
with Eqs.~(\ref{min_max.b}), (\ref{dq_dp.a}) and (\ref{dq_dp.b}),
satisfies
\begin{equation}
  \int_{0}^{t_b} dt' V(q_1(t')) = \left( t_b + \frac{( m
      \sigma^2/\hbar )^2}{t_b} \right) V(q_1(t_b)) \, .
\label{b_general}
\end{equation}
In the limit $t \gg m \sigma^2 / \hbar$, or equivalently $l \gg p_0
\sigma^2 / \hbar$, Eq.~(\ref{b_general}) simplifies to
\begin{equation}
  \int_{q_0}^{b} dq V(q) = |b - q_0| V(b) \, .
\label{b}
\end{equation}
Equation~(\ref{b}) has a simple geometrical interpretation as depicted
in Fig.~\ref{fig-1}: the area under the curve $V(q)$ from $q_0$ to $b$
equals the area of a rectangle with the base $|b - q_0|$ and height
$V(b)$. It is clear from this construction that the LE exhibits a
maximum when the classical unperturbed particle leaves the
perturbation region.

The time instant $t = t_b$, at which $M(t)$ exhibits a maximum, has
the following physical significance. According to
Eq.~(\ref{min_max.b}), and in the limit $l \gg p_0 \sigma^2 / \hbar$,
it satisfies $\Delta q(t_b) = \Delta p(t_b) t_b / m$. This means that
if the perturbed wave packet, $\Psi_2(q; t_b)$, is propagated backward
through time $-t_b$ under the unperturbed (free-particle) Hamiltonian
$\hat{H}_1$ then the resulting wave packet is centered at $q_0$. In
other words, the states $| \Psi_0 \rangle$ and $| e^{i \hat{H}_1 t_b /
  \hbar} e^{-i \hat{H}_2 t_b / \hbar} | \Psi_0 \rangle$ have the same
expectation values of the position operator (while generally different
expectation values of the momentum operator).

Finally, at long times, $t \gg m l / p_0$, one has $\Delta q =
-(m/p_0^2) \int_{-\infty}^{+\infty} dq V(q)$ and $\Delta p = 0$. Thus,
in accordance with Eq.~(\ref{le_free}), the LE saturates at the value
\begin{equation}
  M(\infty) = \exp \left[ -\frac{1}{8}
    \left( \frac{1}{\sigma E} \int_{-\infty}^{+\infty} dq V(q) \right)^2 \right] .
\label{le_satur}
\end{equation}
For the case of a perturbation potential given by Eq.~(\ref{V_Gauss})
the saturation value is 
\begin{equation}
  M(\infty) = \exp \left[ -(\pi/8) (\gamma/\sigma)^2 (V_0/E)^2 \right] \,.
\label{le_satur_2}
\end{equation}
In Fig.~\ref{fig-2} the LE saturation value, $M(\infty)$, calculated
in accordance with Eq.~(\ref{le_satur_2}) is shown by a horizontal
dashed line. We note that the quantitative agreement between the LE
saturation value obtained by numerically solving the Schr{\"o}dinger
equation and the saturation value given by Eqs.~(\ref{le_satur}) and
(\ref{le_satur_2}) improves as the perturbation strength $V_0$ is
reduced. This is consistent with the fact that Eqs.~(\ref{le_satur})
and (\ref{le_satur_2}) are valid only in the leading order of $V_0/E$.

\bigskip

As it will become clear from the following sections the observed
non-monotonicity of $M(t)$ is a generic feature of the LE decay in
one-dimensional (and, more generally, quasi-one-dimensional)
systems. We now turn our attention to closed systems and address the
envelope of $M(t)$. The analysis presented here is valid for short
times, for which the wave function stays localized and can be well
approximated by a Gaussian wave packet. In other words, we restrict
our discussion to $t \ll t_\mathrm{E}$ with $t_\mathrm{E}$ being the
Ehrenfest time.

\subsection{Particle on a ring}
\label{section_ring}

As the fist example of a closed one-dimensional system we consider a
quantum particle moving on a ring. In this set-up the wave function of
the unperturbed system, $\Psi_1(q; t)$, is defined on an interval $0
\leq q \leq L$, and satisfies the time-dependent Schr\"odinger
equation with the Hamiltonian $\hat{H}_1 = \hat{p}^2/2m$ and periodic
boundary condition $\Psi_1(L;t) = \Psi_1(0;t)$. The Hamiltonian of the
perturbed system is $\hat{H}_2 = \hat{H}_1 + V(q)$, where $V(q)$ is
defined on the same $q$-interval with $V(L) = V(0)$. As before, we
assume $V(q)$ to be localized in a small region of size $\gamma$
around a single extremum point $a$, see Fig.~\ref{fig-1}. The wave
function of the perturbed system, $\Psi_2(q;t)$, satisfies the same
periodic boundary conditions, $\Psi_2(L;t) = \Psi_2(0;t)$. Finally, as
in Section~\ref{section_free}, we assume the validity of conditions
given by Eqs.~(\ref{condition-1}--\ref{condition-3}).

As stated above, we only consider the wave packet dynamics for times
shorter than the Ehrenfest time, $t \ll t_{\mathrm{E}}$. The latter is
the time that it takes for the wave packet dispersion, initially equal
to $\sigma$, to become comparable to the size of the system, $L$.  (We
note, that Eq.~(\ref{condition-1}) implies $\sigma \ll L$.) According
to the uncertainty principle the velocity uncertainty of the particle
is $\Delta v \sim \hbar / m\sigma$. Thus, for the case of the free
motion on a ring, the Ehrenfest time can be estimated as
$t_{\mathrm{E}} \sim L/\Delta v \sim m \sigma L / \hbar$.

\begin{figure}[h]
\centerline{\epsfig{figure=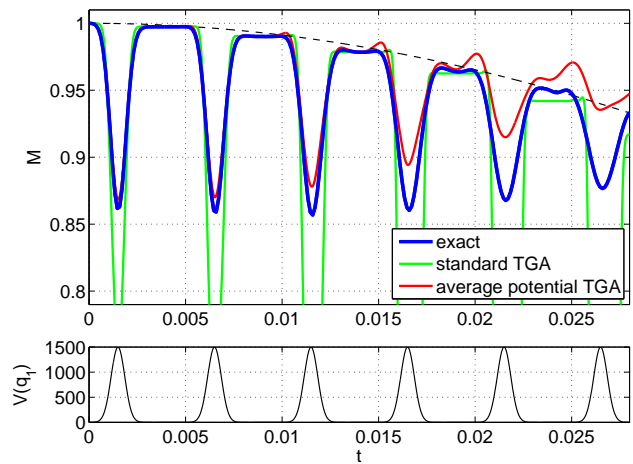,width=3.3in}}
\caption{(Color online) Top figure: The LE, $M(t)$, for a particle on
  a ring with the perturbation potential $V(q)$ given by
  Eq.~(\ref{V_Gauss}). The dashed line represents the envelope decay,
  $\overline{M}(t)$, calculated in accordance with
  Eqs.~(\ref{envelope}) and (\ref{tau_ring_gauss}).  Bottom figure:
  Perturbation potential as seen by the unperturbed particle,
  i.e. $V(q_1(t))$. The system is characterized by $L = 1$, $p_0 =
  200$, $\sigma = 0.1$, $a-q_0 = 0.3$, $\gamma = 0.1$, and $V_0 =
  1500$.}
\label{fig-3}
\end{figure}

Since the dispersion of the wave packet and the extent of the
perturbation potential are small compared to the ring circumference
$L$, the problem in question is effectively equivalent to that of a
free unbounded motion in the presence of a periodic perturbation
potential $\sum_{n=-\infty}^{+\infty} V(q+nL)$. Therefore, according
to Section~\ref{section_free}, we expect the LE curve to exhibit a
sequence of minima at times $t_{\min,k} = m (a-q_0 + kL) / p_0$, with
$k \in \mathbb{N}_0$, at which the unperturbed classical particle
traveling a constant velocity $p_0/m$ passes the extrema of the
periodic perturbation potential. (For concreteness but without loss of
generality, we consider the case $0 < q_0 < a <L$ and $p_0 > 0$.)
Indeed, such a non-monotonous time decay of the LE is found in perfect
agreement with the results of the fully numerical (exact) and
semianalytical (TGA) solutions of the problem, see
Fig.~\ref{fig-3}. We also note here that, just as in Fig.~\ref{fig-2},
$M(t)$ in Fig.~\ref{fig-3} exhibits a sequence of local maxima at
times $t_{\max,k}^+ \simeq m (a-q_0 +\gamma + kL) / p_0$ and
$t_{\max,k+1}^- \simeq m \big(a-q_0 -\gamma + (k+1)L\big) / p_0$, with
$k \in \mathbb{N}_0$. These time instants are solutions of
Eq.~(\ref{b_general}) for $t_b$ with $V(q)$ replaced by the periodic
potential $\sum_{n=-\infty}^{+\infty} V(q+nL)$.

In order to find the envelope function $\overline{M}(t)$ of the LE
decay curve we consider $M$ at times $t_n = n T$, where $n \in
\mathbb{N}$, and $T = mL/p_0$ is the period of the unperturbed
motion. According to Eqs.~(\ref{le_free}) and (\ref{dq_dp.a}) we have
$M(t_n) = \exp(-\Delta q^2/ 2\sigma^2)$ with $\Delta q = -(t_n / p_0L)
\int_0^L dq V(q)$. This leads to
\begin{equation}
  \overline{M}(t) = \exp \left[-(t/\tau)^2\right]
\label{envelope}
\end{equation}
with $\tau = \tau_{\mathrm{ring}}$ and
\begin{equation}
  \tau_{\mathrm{ring}} = \sqrt{2} |p_0|\sigma L \left| \int_0^{L} dq V(q) \right|^{-1} \! .
\label{tau_ring}
\end{equation}
For the case of a Gaussian perturbation potential given by
Eq.~(\ref{V_Gauss}) the expression for the decay time reduces to
\begin{equation}
\tau_{\mathrm{ring}} = T \sqrt{\frac{8}{\pi}} \frac{\sigma E}{\gamma |V_0|} \, ,
\label{tau_ring_gauss}
\end{equation}
where $E=p_0^2/2m$ is the total energy of the system.

The dashed curve in Fig.~\ref{fig-3} is the envelope function
$\overline{M}(t)$ calculated from Eqs.~(\ref{envelope}) and
(\ref{tau_ring_gauss}) for the system specified by $L = 1$, $p_0 =
200$, $\sigma = 0.1$, $a-q_0 = 0.3$, $\gamma = 0.1$, and $V_0 =
1500$. It is evident that $\overline{M}(t)$ accurately describes the
envelope of the exact LE decay represented by the thick blue curve. It
is also interesting to compare the quality of the approximate
solutions.  The standard TGA (green curve) reasonably well
approximates $M(t)$ around its maxima, while completely failing to
capture the function close to its minima. On the other hand, the
average potential TGA (red curve) well describes both maxima and
minima. This approximation however turns out to be only limited to
short times, up to 3 full cycles of the particle around the ring for
the particular set of parameters, after which the approximation
becomes unstable.

\subsection{Particle in a well}
\label{section_well}

\begin{figure}[h]
\centerline{\epsfig{figure=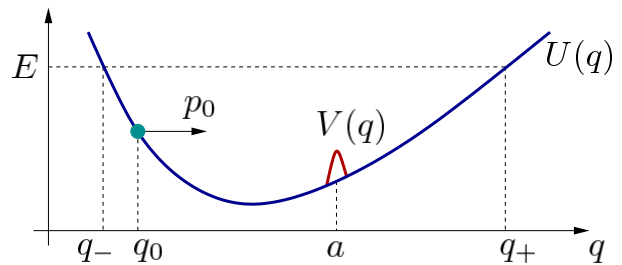,width=3.2in}}
\caption{(Color online) A quantum particle in a one-dimensional
  potential well. The unperturbed potential is given by $U(q)$, and
  the perturbed potential by $U(q)+V(q)$. The total energy $E$ of the
  system is such that both unperturbed and perturbed classical motions
  have the same turning points $q_-$ and $q_+$.}
\label{fig-4}
\end{figure}

We now consider a quantum particle trapped inside a one-dimensional
potential well, so that $\hat{H}_1 = \hat{p}^2/2m + U(\hat{q})$ and
$\hat{H}_2 = \hat{p}^2/2m + U(\hat{q}) + V(\hat{q})$, where $U(q)$
specifies the potential well, and $V(q)$ represents a perturbation
potential localized around $q=a$. As before, we assume validity of the
conditions given by Eqs.~(\ref{condition-1}--\ref{condition-3}) with
the total energy of the classical particle $E = p_0^2/2m + U(q_0)$. We
also assume $E > U(a)+V(a)$, so that both unperturbed and perturbed
classical motions have the same turning points $q_-$ and $q_+$, see
Fig.~\ref{fig-4}.

As we will see below the LE decay for the system under consideration
is non-monotonous. Deep local mimima surrounded by (generally less
pronounced) local maxima occur when the classical unperturbed particle
traverses the perturbation region. The physics underlying this
non-monotonicity is essentially the same as in the cases of a free
particle and a particle on a ring, see Sections~\ref{section_free} and
\ref{section_ring}, so we directly proceed to the analysis of the
envelope function $\overline{M}(t)$ of the LE decay $M(t)$.

We begin by considering the period $T+\Delta T$ of the classical
oscillatory motion of the perturbed system:
\begin{align}
  T+\Delta T &= \sqrt{2m} \int_{q_-}^{q_+}
  \frac{dq}{\sqrt{E-U(q)-V(q)}} \nonumber\\ 
  &\simeq T + \sqrt{\frac{m}{2}} \int_{q_-}^{q_+} dq \, V(q) \big( E-U(q)
  \big)^{-\frac{3}{2}} ,
\label{period_perturbed}
\end{align}
where
\begin{equation}
  T = \sqrt{2m} \int_{q_-}^{q_+} \frac{dq}{\sqrt{E-U(q)}}
\label{period}
\end{equation}
is the period the unperturbed classical
motion. Equation~(\ref{period_perturbed}) holds to the second order in
$V/(E-U)$. Then, assuming that $\big(E-U(q)\big)$ is approximately
constant in a $\gamma$-interval about $q = a$, we obtain
\begin{equation}
  \Delta T = \sqrt{\frac{m}{2}} \big( E-U(a)\big)^{-\frac{3}{2}} \int_{-\infty}^{+\infty} dq V(q) \,.
\label{period_change}
\end{equation}
Here we note that if the perturbation potential satisfies $\int dq
V(q) = 0$ then the expansion in Eq.~(\ref{period_perturbed}) has to be
terminated at the next order resulting in $\Delta T \sim \int dq
V^2(q)$.

The envelope function $\overline{M}(t)$ of the LE decay curve can be
calculated using the approximation given by
Eq.~(\ref{le_same_alpha}). To this end we consider the distance
between the unperturbed and perturbed trajectories in phase space,
$\Delta q = q_2-q_1$ and $\Delta p = p_2-p_1$, at times $t_n = n T$
with $n \in \mathbb{N}$. Since $T$ is the period of the unperturbed
systems we have $q_1(t_n) = q_0$ and $p_1(t_n) = p_0$. The trajectory of
the perturbed system however is ``delayed'' by the time $n\Delta T$
with respect to the unperturbed trajectory. This, for small $\Delta
T$, can be written as $q_2(t_n) = q_1(t_n-n\Delta T)$ and $p_2(t_n) =
p_1(t_n-n\Delta T)$. Expanding these equations to the leading order in
$\Delta T/T$ we obtain
\begin{align}
  &\Delta q(t_n) = -\frac{p_0}{m} \frac{\Delta T}{T} t_n \,, \label{dq_dp_general_well.a}\\
  &\Delta p(t_n) = U'(q_0) \frac{\Delta T}{T} t_n \,, \label{dq_dp_general_well.b}
\end{align}
where the prime denotes the derivative. Then, a substitution of
Eqs.~(\ref{dq_dp_general_well.a}) and (\ref{dq_dp_general_well.b})
into Eq.~(\ref{le_same_alpha}) yields
\begin{equation}
  M(t_n) = \exp \left[ -\chi(t_n) \left( \frac{\Delta T}{T} t_n \right)^2 \right]
\label{envelope_general_well}
\end{equation}
with
\begin{equation}
  \chi = \frac{1}{\Re\alpha_1} \left[
    \left(|\alpha_1| \frac{p_1}{m}\right)^2 - \Im\alpha_1 \frac{p_1
      U'(q_1)}{m \hbar}  + \left( \frac{U'(q_1)}{2\hbar}
    \right)^2 \right] \,,
\label{chi_def}
\end{equation}
keeping in mind that $q_1 = q_1(t)$, $p_1 = p_1(t)$, $\alpha_1 =
\alpha_1(t)$, and $q_1(t_n) = q_0$, $p_1(t_n) = p_0$.

It is now interesting to observe that $\chi$ is a constant of the
motion defined by Eqs.~(\ref{tga.a}--\ref{tga.c}) with $j=1$ and
$W_1(q) = U(q)$. Indeed, it is straightforward to verify that
\begin{equation}
  \frac{d \chi}{dt} = 0 \,.
\label{dchi_dt}
\end{equation}
Then, a replacement of $\chi(t_n)$ in Eq.~(\ref{envelope_general_well})
by its value at time $t=0$,
\begin{equation}
  \chi = \frac{1}{2} \left( \frac{p_0}{m\sigma} \right)^2 
  + \frac{1}{2} \left( \frac{\sigma U'(q_0)}{\hbar} \right)^2 \,,
\label{chi_t0}
\end{equation}
yields the Gaussian LE envelope $\overline{M}(t)$ given by
Eq.~(\ref{envelope}) with the decay time $\tau = \tau_{\mathrm{well}}$
and
\begin{equation}
  \tau_{\mathrm{well}} = \frac{T}{\Delta T} \, \chi^{-\frac{1}{2}} = 
  \sqrt{2} \frac{T}{\Delta T} \left[ \left( \frac{p_0}{m\sigma}\right)^2 
    + \left( \frac{\sigma U'(q_0)}{\hbar} \right) ^2 \right]^{-\frac{1}{2}} .
\label{tau_well}
\end{equation}

Below we consider some particular potentials $U(q)$ and compare the
analytical prediction given by Eq.~(\ref{tau_well}) with the
numerical, full quantum-mechanical solution of the LE decay problem.

\subsubsection{Harmonic oscillator}
\label{section_harmonic}

As our first example we considering motion of a particle in a harmonic
potential well for which $U(q)= m \omega^2 q^2/2$. In this case the
time evolution of the unperturbed system is governed by
\cite{Tan07Introduction}
\begin{align}
  &q_1 = q_0 \cos \omega t + \frac{p_0}{m\omega} \sin \omega t \, , \label{wp_1_harmonic.a} \\
  &p_1 = p_0 \cos \omega t - m \omega q_0 \sin \omega t \, , \label{wp_1_harmonic.b} \\
  &\alpha_1 = \left( \frac{m \omega}{2 \hbar} \right) \frac{\hbar \cos
    \omega t + i m \omega \sigma^2 \sin \omega t}{i \hbar \sin \omega
    t + m \omega \sigma^2 \cos \omega t} \, . \label{wp_1_harmonic.c}
\end{align}
Equations~(\ref{wp_time}) and
(\ref{wp_1_harmonic.a}--\ref{wp_1_harmonic.c}) give the exact quantum
dynamics of the system with the Hamiltonian $\hat{H}_1$ and initial
state of Eq.~(\ref{init_wp}). As before, the time dependent parameters
$q_2$, $p_2$, and $\alpha_2$ characterizing the perturbed wave packet
are calculated using the standard TGA,
Eqs.~(\ref{tga.a}--\ref{tga.c}), or the average potential TGA,
Eqs.~(\ref{avr_tga.a}--\ref{avr_pot}), with $W_2(q) =
U(q)+V(q)$. Then, the TGA approximation of the LE is computed from
Eq.~(\ref{le_general}), and the envelope function, $\overline{M}(t)$,
from Eqs.~(\ref{envelope}) and (\ref{tau_well}) with $T = 2\pi/\omega$
and $\Delta T$ given by Eq.~(\ref{period_change}).

\begin{figure}[h]
\centerline{\epsfig{figure=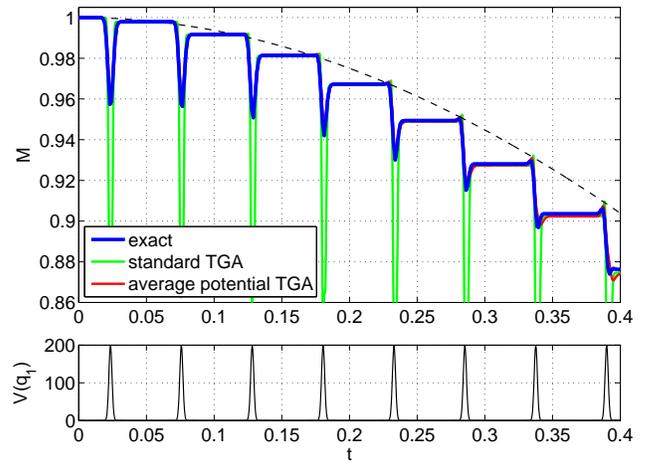,width=3.3in}}
\caption{(Color online) Top figure: the LE, $M(t)$, for a particle in
  a parabolic well with the perturbation potential $V(q)$ given by
  Eq.~(\ref{V_Gauss}). The dashed line represents the envelope decay,
  $\overline{M}(t)$, calculated in accordance with
  Eqs.~(\ref{envelope}) and (\ref{tau_well}). Bottom figure: the
  perturbation potential as seen by the unperturbed particle,
  i.e. $V(q_1(t))$. The system is characterized by $\omega = 60$, $q_0
  = -1$, $p_0 = 10$, $\sigma = 0.1$, $a = 0$, $\gamma = 0.1$, and $V_0
  = 200$. Note that the LE decay obtained by the average potential TGA
  (red curve) is essentially indistinguishable from the exact result
  (blue thick curve) for most of the time range.}
\label{fig-5}
\end{figure}

Figure \ref{fig-5} shows the time decay of the LE for a quantum
particle in the harmonic potential well. The unperturbed system is
defined by $\omega = 60$, $q_0 = -1$, $p_0 = 10$, $\sigma = 0.1$. The
perturbation potential is given by Eq.~(\ref{V_Gauss}) with $a = 0$,
$\gamma = 0.1$, and $V_0 = 200$. As before, the thick blue curve
represents $M(t)$ obtained by numerically solving the time dependent
Schr\"odinger equation for the perturbed system. (The exact time
evolution of the unperturbed system is given by Eqs.~(\ref{wp_time})
and (\ref{wp_1_harmonic.a}--\ref{wp_1_harmonic.c}).) The green curve
corresponds to the standard TGA for the perturbed system, i.e.,
Eqs.~(\ref{tga.a}--\ref{tga.c}) with $W_2(q) = m \omega^2 q^2/2 +
V(q)$. It is clear from the figure that the standard TGA fails to
reproduce the minima of the exact LE curve. On the other hand, the
average potential TGA result, computed according to
Eqs.~(\ref{avr_tga.a}--\ref{avr_pot}) and represented by the red
curve, provides an excellent approximation of the exact result. The
dashed line shows the envelope decay, $\overline{M}(t)$, predicted by
Eqs.~(\ref{envelope}) and (\ref{tau_well}). A reasonable agreement
between the theoretical and numerical results is apparent.

\subsubsection{Anharmonic oscillator}
\label{section_anharmonic}

We now consider, as our unperturbed system, a quantum particle moving
in an anharmonic potential well of the form $U(q) = m \omega^2 q^2/2 +
\epsilon q^3$. Here, the computational procedure of the LE decay is
essentially the same as in Section~\ref{section_harmonic} with the
only difference that the oscillation period $T$ is obtained by
numerically evaluating the integral in Eq.~(\ref{period}).

\begin{figure}[h]
\centerline{\epsfig{figure=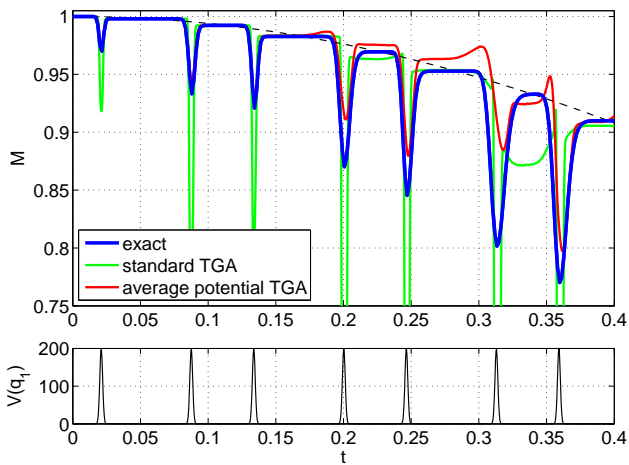,width=3.3in}}
\caption{(Color online) Top figure: the LE, $M(t)$, for a particle in
  an anharmonic well with the perturbation potential $V(q)$ given by
  Eq.~(\ref{V_Gauss}). The dashed line represents the envelope decay,
  $\overline{M}(t)$, calculated in accordance with
  Eqs.~(\ref{envelope}) and (\ref{tau_well}). Bottom figure: the
  perturbation potential as seen by the unperturbed particle,
  i.e. $V(q_1(t))$. The system is characterized by $\omega = 60$,
  $\epsilon = -400$, $q_0 = -1$, $p_0 = 10$, $\sigma = 0.1$, $a = 0$,
  $\gamma = 0.1$, and $V_0 = 200$.}
\label{fig-6}
\end{figure}

Figure \ref{fig-6} shows the time decay of the LE for a quantum
particle in the anharmonic potential well. The unperturbed system is
defined by $\omega = 60$, $\epsilon = -400$, $q_0 = -1$, $p_0 = 10$,
$\sigma = 0.1$. The perturbation potential is given by
Eq.~(\ref{V_Gauss}) with $a = 0$, $\gamma = 0.1$, and $V_0 = 200$. The
thick blue curve represents the exact LE decay. The green curve
corresponds to the standard TGA for the unperturbed and perturbed
systems, i.e., Eqs.~(\ref{tga.a}--\ref{tga.c}) with $W_1(q) = U(q)$
and $W_2(q) = U(q) + V(q)$. Once again, the standard TGA fails to
reproduce the minima of the exact LE curve. The red curve shows the
result of the average potential TGA, i.e.,
Eqs.~(\ref{avr_tga.a}--\ref{avr_pot}) with $W_1(q) = U(q)$ and $W_2(q)
= U(q) + V(q)$. It well approximates the exact LE decay curve for up
to two full oscillations of the unperturbed system.  The dashed line
shows the envelope decay, $\overline{M}(t)$, predicted by
Eqs.~(\ref{envelope}) and (\ref{tau_well}). As in
Section~\ref{section_harmonic}, one observes a reasonable agreement
between the theoretical and numerical results.

\section{Beyond one dimension}
\label{section_2d}

In this section we argue that the reported non-monotonicity of the
short-time decay of the LE is pertinent mainly to one-dimensional (or,
more generally, quasi-one-dimensional) systems. The disappearance of
the non-monotonicity of the LE decay in systems with two or higher
number of dimensions can be understood as follows. In one-dimensional
systems an initially Gaussian wave packet traverses the perturbation
region and approximately preserves its Gaussian shape provided the
strength of the perturbation potential is small compared to the energy
of the particle. In this case the TGA provides a natural basis for the
LE analysis and the theory developed in Section~\ref{section_1d}
holds. In higher number of dimensions however it is often
energetically preferable for the wave packet to split in parts and
circumvent the perturbation region. The wave packet deforms to avoid
the perturbation region and can no longer be approximated by a
Gaussian. The overlap between the unperturbed (approximately Gaussian)
and perturbed (substantially non-Gaussian) wave packets decays rapidly
and monotonically during the time that the system interacts with the
perturbation.

To illustrate this point we consider the time decay of the LE for a
free quantum particle in two dimensions under the action of a Gaussian
perturbation potential. The initial state of the particle is given by
\begin{align}
  \Psi_0(x,y) = \frac{1}{\sqrt{\pi} \sigma} \exp \bigg(
  &\frac{i}{\hbar} p_0(x-x_0) \nonumber\\ & -
  \frac{(x-x_0)^2+(y-y_0)^2}{2\sigma^2} \bigg) \,.
\label{init_wp_2d}
\end{align}
Here, $(x_0,y_0)$ is the center of the wave packet, $p_0$ the
magnitude of the momentum pointing along the $x$-axis, and $\sigma$
its dispersion. The perturbation potential is defined as
\begin{equation}
  V(x,y) = V_0 \exp\left( -\frac{(x-a)^2}{\gamma^2} - \frac{y^2}{\delta^2} \right)
\label{V_Gauss_2d}
\end{equation}
with $\gamma$ and $\delta$ quantifying the extent of the perturbation
region in the $x$- and $y$-directions respectively. As in
Section~\ref{section_1d} we assume the validity of the conditions
given by Eqs.~(\ref{condition-1}--\ref{condition-3}).

\begin{figure}[h]
\centerline{\epsfig{figure=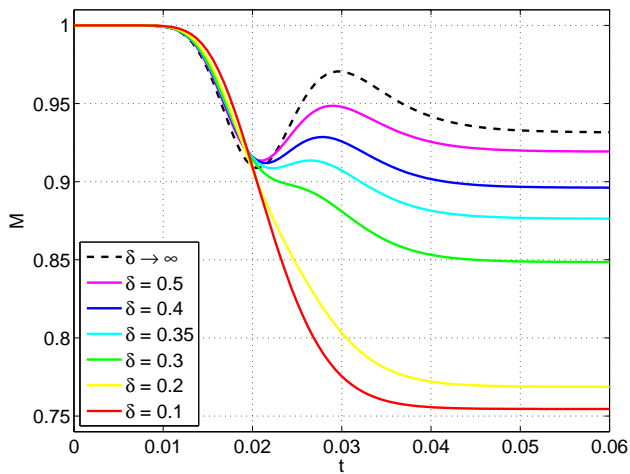,width=3.3in}}
\caption{(Color online) The LE decay for a free particle in two
  dimensions under the action of the perturbation potential given by
  Eq.~(\ref{V_Gauss_2d}). The system is characterized by $p_0 = 50$,
  $\sigma = 0.1$, $a-q_0 = 1$, $V_0 = 150$, and $\gamma = 0.3$. The
  $M(t)$ curves correspond to seven different values of the $y$-extent
  of the perturbation region: from bottom to top $\delta = 0.1$,
  $0.2$, $0.3$, $0.35$, $0.4$, $0.5$, and $\infty$.}
\label{fig-7}
\end{figure}

Figure~\ref{fig-7} shows the time decay of the LE, $M(t)$, obtained by
numerically solving the time-dependent Schr\"odinger equation with the
initial state given by Eq.~(\ref{init_wp_2d}) and the Hamiltonian
operators $\hat{H}_1 = (-\hbar^2/2m)(\partial^2 / \partial x^2
+ \partial^2 / \partial y^2)$ and $\hat{H}_2 = \hat{H}_1 +
V(\hat{x},\hat{y})$, where $V(x,y)$ is defined by
Eq.~(\ref{V_Gauss_2d}). The system parameters are $p_0 = 50$, $\sigma
= 0.1$, $a-q_0 = 1$, $V_0 = 150$, and $\gamma = 0.3$. Six different
values of the $y$-axis extent of the perturbation region are
considered, $\delta = 0.1$, $0.2$, $0.3$, $0.35$, $0.4$, $0.5$, as
well as the case $\delta \rightarrow \infty$ corresponding to the
one-dimensional problem analyzed in Section~\ref{section_free}. It is
clear from the figure that in quasi-one-dimensional systems, i.e.,
when $\sigma \ll \delta$ so that the $y$-component of the classical
force exerted on the particle by the perturbation potential is
negligible, $M(t)$ exhibits minima and maxima as the particle
traverses the perturbation region. On the other hand, in substantially
two-dimensional systems, i.e., for $\delta \lesssim \sigma$, a fast
monotonous decay of the LE is observed.

\begin{figure}[h]
\centerline{\epsfig{figure=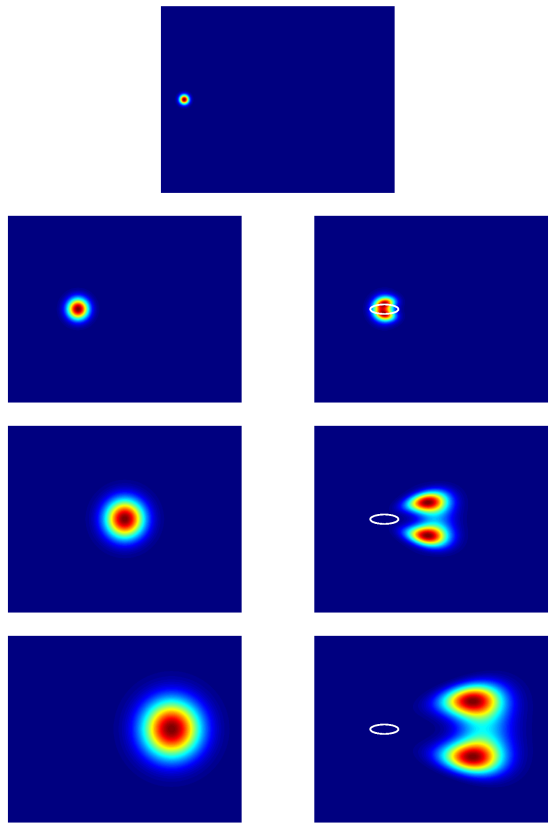,width=3.2in}}
\caption{(Color online) The time evolution of the unperturbed (left
  column) and perturbed (right column) wave function. The top snapshot
  shows the initial state. The white ellipse in the snapshots of the
  right column is centered at the position of the perturbation
  maximum, and has its major and minor semi-axes equal to $\gamma =
  0.3$ and $\delta = 0.1$ respectively. The system is characterized by
  $p_0 = 50$, $\sigma = 0.1$, $a-q_0 = 1$, $V_0 = 150$, $\gamma =
  0.3$, $\delta = 0.1$, and exhibits a LE decay represented by the red
  (most bottom) curve in Fig.~\ref{fig-7}.}
\label{fig-8}
\end{figure}

In order to support our qualitative explanation for the disappearance
of the non-monotonicity in the LE decay for small values of $\delta$
we depict in Fig.~\ref{fig-8} the time evolution of the unperturbed
and perturbed wave functions for the system characterized by $p_0 =
50$, $\sigma = 0.1$, $a-q_0 = 1$, $V_0 = 150$, $\gamma = 0.3$, $\delta
= 0.1$. This set of parameters corresponds to the LE decay represented
by the red (most bottom) curve in Fig.~\ref{fig-7}. Here, a snapshot
at the top shows the probability distribution of the initial
state. The left column shows the probability distribution of the
unperturbed wave packet at regular time intervals. The corresponding
snapshots in the right column give the probability distribution of the
perturbed wave function. A white ellipse in the right column shows the
characteristic extent of the perturbation potential: the ellipse is
centered at the maximum of the Gaussian potential, and its major and
minor semi-axes equal to $\gamma = 0.3$ and $\delta = 0.1$
respectively. The figure demonstrates how the probability distribution
of the perturbed wave function changes from a simple Gaussian to a
double-peak function as the quantum particle circumvents the
perturbation region. Clearly, this process can not be adequately
described by the TGA. It is also natural to expect the LE overlap to
decay monotonically and faster than in the limit of large $\delta$, in
which the perturbed wave function does not significantly change its
functional form and only gets displaced in phase space with respect to
the unperturbed wave packet.

The qualitative change of the LE decay shown in Fig.~\ref{fig-7} can
be linked to an effective decrease of the Ehrenfest time
$t_\mathrm{E}$ of the perturbed systems as the perturbation size
$\delta$ decreases: the observed splitting of the wave packet
indicates that $t_\mathrm{E} \sim m (a-q_0) / p_0$ for small values
of $\delta$.

It is interesting to note that the phenomenon of wave packet splitting
is observed already for relatively weak perturbation potentials. For
example, the right column in Fig.~\ref{fig-8} corresponds to the
energy of the classical particle, $p_0^2/2$, being more than 8 times
higher than the strength of the potential, $V_0$. At this point, it is
not clear whether the splitting can be adequately explained in terms
of deflection of the corresponding classical trajectories or whether
quantum interference effects play a crucial role. It can be speculated
that the wave packet splitting is intimately related to the phenomenon
of coherent electron flow branching in two-dimensional quantum wells
with weak disordered background potentials \cite{TLW+01Coherent,
  TWH03Imaging, vanicek_heller_03}. However, a proper investigation is
yet to be performed.

As a final remark we make the following counter-intuitive
observation. By increasing the parameter $\delta$, while keeping $V_0$
fixed, one increases the overall presence of the perturbation, in the
sense that the integral $\int dx dy \, V(x,y)$ becomes
larger. However, at the same time, the larger $\delta$ the higher the
LE saturation value, see Fig~\ref{fig-7}, and therefore the smaller
the effect of the perturbation on the time evolution of the wave
function. The reason for that is that it is energetically preferable
for a wave function to avoid small perturbation regions by changing
its shape. This results in small values of the LE overlap. On the
other hand, spatially extended perturbation regions are traversed by
the wave function without any significant shape change resulting in
large values of the LE overlap.

\section{Conclusions}
\label{section_conclusions}

We have studied, analytically and numerically, the decay of the
Loschmidt echo (LE) from local Hamiltonian perturbations for times
short compared to the Ehrenfest time. During those times the quantum
states of the unperturbed and perturbed systems can be efficiently
approximated by localized Gaussian wave packets. The latter evolve in
accordance with equations of the thawed Gaussian approximation
(TGA). We have analyzed these equations and deduced several important
properties of the LE decay in quasi-one-dimensional systems.

Firstly, we have shown that the LE is generally a non-monotonous
function of time. More specifically, $M(t)$ exhibits strongly
pronounced minima and maxima at the instants of time when the
corresponding classical particle traverses the perturbation
region. The minima of $M(t)$ are reached when the particle passes the
peak (or bottom) of the perturbation potential, whereas the maxima
generally occur at the times the particle enters or exits the
perturbation region. We have also demonstrated that while the observed
non-monotonicity of the LE decay is generic in one-dimensional systems
the short-time decay can be monotonous in systems with higher number
of dimensions.

Secondly, using the TGA we have analyzed the envelope decay of the LE
and shown it to be well approximated by a Gaussian, $\overline{M}(t) =
\exp\left[ -(t/\tau)^2\right]$. The decay time $\tau$ is expressible
in terms of parameters of the initial state, system's Hamiltonian, and
perturbation potential. We have given explicit formulas for the cases
of a particle on a ring, Eq.~(\ref{tau_ring}), and a particle inside a
smooth potential well of an arbitrary shape, Eq.~(\ref{tau_well}). The
analytical formulas have been given convincing numerical support.

All results presented in this paper pertain to the short-time decay of
the LE in ``clean'' quantum systems and assume no averaging over
initial states or Hamiltonian perturbations. In this respect, the
findings may prove valuable in echo experiments which imply no or
minimal averaging intrinsically. Thus, for example, in echo
experiments with ultra-cold atoms one could try to use an
experimentally observed decay function $M(t)$ to reconstruct the
localized perturbation potential.

\section*{Acknowledgments}

This work in its early stage was supported by EPSRC under Grant
No.~EP/E024629/1. The author would like to thank Philippe Jacquod and
Klaus Richter for reading and helping to improve the manuscript.

\appendix
\section{Thawed Gaussian approximations}
\label{section_TGA}

The celebrated thawed Gaussian approximation (TGA), originally
introduced by Heller \cite{Hel75Time}, is a semiclassical technique
for propagation of localized Gaussian wave packets in time without
having to solve the full Schr\"odinger equation. The approximation is
based on the idea that, at least for short times, the expectation
values of the position and momentum operators, $q_j$ and $p_j$
respectively, evolve according to the classical, Hamilton equations of
motion. Thus, in the TGA the evolving wave packet is ``guided'' by a
single real phase-space trajectory described by the Hamilton
equations. The TGA is one of the simplest approximations to the full
Van Vleck-Gutzwiller semiclassical propagator \cite{Gut90Chaos}.

Below we begin with a brief discussion of the standard TGA following
Ref.~\cite{Hel75Time}. A comprehensive review of the subject can be
found in Ref.~\cite{Hel06Guided}. We then introduce an extended
version of the approximation -- the average potential TGA -- that
proves more reliable for potentials with rapid spatial variations.

The central assumption of the TGA is that an initially Gaussian wave
packet preserves its Gaussian form, $\Psi_j(q;t)$ as given by
Eq.~(\ref{wp_time}), throughout the time evolution under a Hamiltonian
$\hat{H}_j = \hat{p}^2/2m + W_j(\hat{q})$. The wave packet is
parametrized by one complex-valued -- $\alpha_j(t)$ -- and three
real-valued -- $q_j(t)$, $p_j(t)$, and $\phi_j(t)$ -- functions, which
are determined by substituting $\Psi_j$ into the Schr\"odinger
equation:
\begin{align}
  \Bigg[ &\left( i\hbar \dot{\alpha}_j - \frac{2\hbar^2}{m} \alpha_j^2
  \right) (q-q_j)^2 + \dot{p}_j (q-q_j) \nonumber \\
  &- 2i\hbar \alpha_j \left( \dot{q}_j - \frac{p_j}{m} \right) (q-q_j)
  + \dot{\phi}_j - p_j\dot{q}_j \nonumber\\ &+\frac{p_j^2}{2m} +
  W_j(q) +\frac{\hbar^2}{m}\alpha_j -
  \frac{i\hbar}{4}\frac{\Re\dot{\alpha}_j}{\Re\alpha_j} \Bigg]
  \Psi_j(q;t) = 0 \, .
\label{se_subs}
\end{align}
Here dots represent differentiation with respect to time. Clearly,
Eq.~(\ref{se_subs}) can not be satisfied for a general potential
$W_j(q)$. However, an approximate solution of Eq.~(\ref{se_subs}) can
be found by expanding the potential $W_j(q)$ into the power series
about $q=q_j$ to the second order in $(q-q_j)$, and then separately
comparing terms of the same order. This procedure yields
Eqs.~(\ref{tga.a}--\ref{tga.c}) together with $\dot{\phi}_j = p_j^2/2m
- W_j(q_j) - (\hbar^2/m)\Re\alpha_j$.

The standard TGA, Eqs.~(\ref{tga.a}--\ref{tga.c}), relies on the
quadratic approximation of the potential $W_j(q)$ about $q=q_j$, and
is therefore limited to potentials varying slowly on the scale given
by the spatial extent of the wave packet. In order to lift this
limitation we modify the TGA method as discussed below.

The problem of finding the ``best'' functions $\alpha_j(t)$, $q_j(t)$,
$p_j(t)$, and $\phi_j(t)$ approximating the time evolution of the wave
packet allows for an alternative approach. If $\Psi_j(q;t)$, given by
Eq.~(\ref{wp_time}), were a true solution of the time-dependent
Schr\"odinger equation with the Hamiltonian $\hat{H}_j = \hat{p}^2/2m
+ W_j(\hat{q})$ then the expectation value of any (generally
time-dependent) operator $\hat{O}_j$ would satisfy
\begin{equation}
  \frac{d}{dt} \langle \Psi_j | \hat{O}_j | \Psi_j \rangle  =
  \frac{i}{\hbar} \langle \Psi_j | [\hat{H}_j, \hat{O}_j] | \Psi_j \rangle 
  + \langle \Psi_j | \frac{\partial \hat{O}_j}{\partial t} | \Psi_j \rangle
\label{expect_evol}
\end{equation}
with $[\cdot,\cdot]$ denoting the commutator. For a general potential
$W_j$, however, $\Psi_j(q;t)$ given by Eq.~(\ref{wp_time}) is not a
true solution of the Schr\"odinger equation, and, therefore,
Eq.~(\ref{expect_evol}) can only be satisfied by four linearly
independent Hermitian operators $\hat{O}_j$. A choice of these four
operators uniquely defines the functions $\alpha_j(t)$, $q_j(t)$, and
$p_j(t)$. (Equation~(\ref{expect_evol}) is unaffected by the global
phase $\phi_j$. This phase however is of no importance to the problem
of the Loschmidt echo decay addressed in this paper.) Although there
is no unique choice of the four independent Hermitian operators
$\hat{O}_j$ we proceed with the seemingly most natural ones:
$\hat{q}$, $\hat{p}$, $\hat{q}^2$, and $\hat{p}^2$. Straightforward
calculations yield Eqs.~(\ref{avr_tga.a}--\ref{avr_pot}) for the time
evolution of $q_j$, $p_j$, and $\alpha_j$. Obviously, the requirement
that $\Psi_j(q;t)$ satisfies Eq.~(\ref{expect_evol}) for $\hat{O}_j =
\hat{q}, \; \hat{p}, \; \hat{q}^2, \; \hat{p}^2$ guaranties the
resulting approximation, which we here term the average potential TGA,
to predict the phase space dynamics of the wave packet's center and
dispersion in the best possible way.

Here two final remarks are in order. Firstly, since
$\widetilde{W}_j'(q_j;\Re\alpha_j) = \langle \Psi_j | W_j'(\hat{q}) |
\Psi_j \rangle$ Eqs.~(\ref{avr_tga.a}) and (\ref{avr_tga.b}) are
nothing but the statement of the Ehrenfest theorem
\cite{Tan07Introduction}. Secondly, it can be readily shown that in
the case of the potential $W_j(q)$ being a second order polynomial in
$q$ Eqs.~(\ref{tga.a}--\ref{tga.c}) are identical with
Eqs.~(\ref{avr_tga.a}--\ref{avr_tga.c}). In this case $\Psi_j(q;t)$
constitutes the true solution of the Schr\"odinger equation.


The idea of exploiting average potentials for propagation of Gaussian
wave packets has been previously discussed in the chemical physics
literature, e.g., see Ref.~\cite{CSS04Properties}. These approaches
however deal with potentials averaged with respect to Gaussian wave
functions with ``frozen'' exponents (such as $\Psi_j(q;t)$ in
Eq.~(\ref{wp_time}) with $\alpha_j(t) = \alpha_0$) and are generally
different from the average potential TGA presented above. It would be
interesting to compare these different methods in terms of their
precision and computational efficiency.


\end{document}